\documentstyle[12pt]{article}
\begin{document}
\tolerance=5000
\def\be{\begin{equation}}
\def\ee{\end{equation}}
\def\bea{\begin{eqnarray}}
\def\eea{\end{eqnarray}}
\def\nn{\nonumber \\}
\def\cF{{\cal F}}
\def\det{{\rm det\,}}
\def\Tr{{\rm Tr\,}}
\def\e{{\rm e}}
\def\etal{{\it et al.}}
\def\erp2{{\rm e}^{2\rho}}
\def\erm2{{\rm e}^{-2\rho}}
\def\er4{{\rm e}^{4\rho}}
\def\etal{{\it et al.}}
\def\gsim{\ ^>\llap{$_\sim$}\ }

\ 

\vskip -2cm

\ \hfill
\begin{minipage}{3.5cm}
NDA-FP-66 \\
June 1999 \\
\end{minipage}

\vfill

\begin{center}
{\Large\bf Running Gauge Coupling and Quark-Antiquark Potential 
in Non-SUSY 
Gauge Theory at Finite Temperature from IIB SG/CFT correspondence}

\vfill

{\sc Shin'ichi NOJIRI}\footnote{\scriptsize 
e-mail: nojiri@cc.nda.ac.jp} and
{\sc Sergei D. ODINTSOV$^{\spadesuit}$}\footnote{\scriptsize 
e-mail: odintsov@mail.tomsknet.ru}

\vfill

{\sl Department of Mathematics and Physics \\
National Defence Academy, 
Hashirimizu Yokosuka 239, JAPAN}

\ 

{\sl $\spadesuit$ 
Tomsk Pedagogical University, 634041 Tomsk, RUSSIA \\
}

\ 

\vfill

{\bf abstract}

\end{center}

We discuss the non-constant dilaton deformed 
${\rm AdS}_5\times{\rm S}_5$ 
solutions of IIB supergravity where AdS sector is described 
by black hole. The investigation of running gauge coupling 
(exponent of dilaton) of non-SUSY gauge theory at finite 
temperature is presented for different regimes (high or 
low $T$, large radius expansion). Running gauge coupling 
shows power-like behavior on temperature with stable fixed point.
The quark-antiquark potential at finite $T$ is found and possibility 
of confinement is established. It is shown that non-constant 
dilaton affects the potential, sometimes reversing its behavior 
if we compare it with the constant dilaton case (${\cal N}=4$ 
super Yang-Mills theory).
Thermodynamics of obtained backgrounds is studied. In particular, 
next-to-leading term to free energy $F$ is evaluated as 
$F=-{\tilde V_3 \over 4\pi^2 }\left(
{N^2 \left(\pi T\right)^4 \over 2} 
+ {5c^2 \over 768 g_{YM}^6 N{\alpha'}^6 
\left(\pi T\right)^4}\right)$. Here $\tilde V_3$ is the volume of 
the space part in the boundary of AdS, $c$ is the parameter 
coming from the non-constant dilaton and $N$ is the number of the 
coincident D3-branes.

\newpage

\section{Introduction.}

One of the unsolved problems in AdS/CFT correspondence \cite{1} 
(for an excellent review, see \cite{AGMOO}) is how to obtain non-SUSY 
gauge theory with typical running coupling as the boundary side. 
The related question is about confinement in such theory. 
It is desirable to answer these questions from supergravity (SG) 
side as it gives strong coupling regime of boundary quantum field 
theory (QFT).

There are different proposals to get running gauge coupling 
in non-SUSY theory: using Type 0 string theory approach 
\cite{2}, deforming ${\cal N}=4$ theory \cite{15} 
(also via AdS orbifolding \cite{17}) or making non-constant dilaton 
deformations of ${\rm AdS}_5\times{\rm S}_5$ vacuum 
in IIB SG \cite{3,4,6,7,8,9,NO}. 
In the last case non-constant dilaton breaks conformal 
invariance and (a part of) supersymmetry of the boundary ${\cal N}=4$ 
super YM theory. (In the presence of axion (RR-scalar), a 
part of supersymmetry may be unbroken but 
dilaton is still non-trivial \cite{14}). Then, exponent 
of dilaton actually describes the running gauge coupling 
with a power-law behavior and UV-stable fixed point. 
Within such picture the indication to the possibility of 
confinement is also found. The features of running and confinement 
depend on the axion \cite{7}, vectors \cite{8}, 
worldvolume scalar \cite{9} or curvature of four-dimensional 
space \cite{NO}.

 From another side, it is also realized that planar 
${\rm AdS}_5$ BH is dual to a thermal state of ${\cal N}=4$ 
super YM theory. The corresponding coupling constant 
dependence has been studied in ref.\cite{GKT,AAT} based on 
earlier study of SG side free energy in ref.\cite{GKP}. 
Spherical AdS BH shows the finite temperature phase transition 
\cite{HP} which may be used to realize the confinement 
in large $N$ theory at low temperatures \cite{witten}.

In this paper, we attempt to combine these two approaches, 
i.e. to find the deformation of IIB SG ${\rm AdS}_5\times{\rm S}_5$ 
background with non-trivial dilaton where AdS sector is described 
by BH (hence, temperature appears). Then, running gauge coupling 
of gauge theory at non-zero temperature is given by exponent of 
dilaton. We present the class of approximate solutions of 
IIB SG\footnote{These solutions presumably describe thermal states 
of non-SUSY gauge theory which descends from ${\cal N}=4$ super YM 
after breaking of SUSY and conformal invariance.} with such 
properties where running coupling shows power-like behavior in the 
temperature (in the expansion on radius). The quark-antiquark 
potential for these solutions is also found and possibility of 
confinement at non-zero temperature is established. Corrections to 
position of horizon (in near horizon regime) and to the temperature 
are calculated. Thermodynamics of obtained solutions is also 
investigated.

The paper is organized as follows. In the next section we present 
the approximate solution of IIB supergravity. It represents dilatonic 
perturbation of zero mass hyperbolic AdS BH. The temperature 
dependence of running gauge coupling (exponent of dilaton) and of 
corresponding beta-function is derived in different regimes. 
Quark-antiquark potential which is repulsive unlike to constant 
dilaton case is analyzed. Section 3 is devoted to the study of 
the same questions for background representing dilatonic deformation 
of (non)planar non-zero mass AdS BH. The temperature dependence 
of running gauge coupling is different from the situation in 
previous section. Confinement is possible as it follows from the 
study of quark-antiquark potential. In section 4, we investigate 
thermodynamic properties of our AdS backgrounds. Free energy, mass 
and entropy are found with account of non-trivial temperature 
corrections due to dilaton. This is compared with the leading 
behaviour of free energy in $N=4$ super Yang-Mills theory. Some 
outlook is given in the last section.

\section{Perturbative solutions of IIB supergravity,
running gauge coupling and potential: zero mass BH case}

We start from the action of dilatonic gravity
in $d+1$ dimensions:
\be
\label{i}
S=-{1 \over 16\pi G}\int d^{d+1}x \sqrt{-G}\left(R - \Lambda 
- \alpha G^{\mu\nu}\partial_\mu \phi \partial_\nu \phi \right)\ .
\ee
In the following, we assume $\lambda^2\equiv -\Lambda$ 
and $\alpha$ to be positive. The action (\ref{i}) contains the 
effective action of type IIB string theory. In the type IIB 
supergravity, which is the low energy effective action of the 
type IIB string theory, we can consider bosonic background where 
anti-self-dual five-form is given by the Freund-Rubin-type ansatz 
and the topology is $M_5\times {\rm S}^5$ with the manifold $M_5$ 
which is asymptotically ${\rm AdS}_5$. If dilaton only depends on 
the coordinates in $M_5$, by integrating five coordinates on 
${\rm S}^5$, we obtain the effective five dimensional theory, which 
corresponds to $d=4$ and $\alpha={1 \over 2}$ case in (\ref{i}). 
This will be the case under consideration in this work.

 From the variation of the action (\ref{i}) with respect to the 
 metric $G^{\mu\nu}$, we obtain\footnote{
The conventions of curvatures are given by
\begin{eqnarray*}
R&=&G^{\mu\nu}R_{\mu\nu} \\
R_{\mu\nu}&=& -\Gamma^\lambda_{\mu\lambda,\kappa}
+ \Gamma^\lambda_{\mu\kappa,\lambda}
- \Gamma^\eta_{\mu\lambda}\Gamma^\lambda_{\kappa\eta}
+ \Gamma^\eta_{\mu\kappa}\Gamma^\lambda_{\lambda\eta} \\
\Gamma^\eta_{\mu\lambda}&=&{1 \over 2}G^{\eta\nu}\left(
G_{\mu\nu,\lambda} + G_{\lambda\nu,\mu} - G_{\mu\lambda,\nu} 
\right)\ .
\end{eqnarray*}
}
\be
\label{iit}
0=R_{\mu\nu}-{1 \over 2}G_{\mu\nu}R + {\Lambda \over 2}G_{\mu\nu}
- \alpha \left(\partial_\mu\phi\partial_\nu\phi 
-{1 \over 2}G_{\mu\nu}G^{\rho\sigma}\partial_\rho \phi
\partial_\sigma \phi \right)
\ee
and from that of dilaton $\phi$
\be
\label{iiit}
0=\partial_\mu\left(\sqrt{-G}G^{\mu\nu}\partial_\nu\phi\right)\ .
\ee
We now assume the $(d+1)$-dimensional metric is given by
\be
\label{ii}
ds^2=-\e^{2\rho}dt^2 + \e^{2\sigma}dr^2 
+ r^2 \sum_{i,j=1}^{d-1}g_{ij}dx^i dx^j\ .
\ee
Here $g_{ij}$ does not depend on $r$ and it is the metric in the 
Einstein manifold, which is defined by
\be
\label{vat}
\hat R_{ij}=kg_{ij}\ .
\ee
Here $\hat R_{ij}$ is Ricci tensor defined by $g_{ij}$ and $k$ is 
a constant, especially $k>0$ for sphere , $k=0$ for Minkowski space 
and $k<0$ for hyperboloid. We also assume $\rho$, $\sigma$ and 
$\phi$ only depend on $r$. Then the equations (\ref{iit}) when 
$\mu=\nu=t$, $\mu=\nu=r$ and $\mu=i$ and $\nu=j$ give, respectively,
\bea
\label{iii}
0&=&{(d-1)k\e^{2\sigma} \over 2r^2} + {(d-1)\sigma' \over r} 
- {(d-1)(d-2) \over 2r^2} \nn
&& + {\lambda^2 \over 2}\e^{2\sigma} 
- {\alpha \over 2}\left(\phi'\right)^2 \\
\label{iv}
0&=&-{(d-1)k\e^{2\sigma} \over 2r^2} + {(d-1)\rho' \over r} 
+ {(d-1)(d-2) \over 2r^2} \nn
&& - {\lambda^2 \over 2}\e^{2\sigma} 
- {\alpha \over 2}\left(\phi'\right)^2 \\
\label{v}
0&=&-{(d-3)k\e^{2\sigma} \over 2r^2} + \rho'' 
+ \left(\rho'\right)^2 - \rho'\sigma' 
+ {(d-2)(d-3) \over 2r^2} \nn
&& - {\lambda^2 \over 2}\e^{2\sigma} 
+ {\alpha \over 2}\left(\phi'\right)^2 \ .
\eea
Here $'\equiv {d \over d r}$. Other components give identities. 
Eq.(\ref{iiit}) has the following form 
\be
\label{vi}
0=\left(r^{d-1}\e^{\rho - \sigma}\phi'\right)'\ ,
\ee
which can be integrated to give 
\be
\label{vii}
r^{d-1}\e^{\rho - \sigma}\phi'=c\ .
\ee
Combining (\ref{iii}) and (\ref{iv}) and substituting 
(\ref{vii}), we obtain
\bea
\label{viii}
0&=&{(d-1)\left(\rho' + \sigma'\right) \over r} 
- {\alpha c^2 \e^{2\sigma - 2\rho} \over r^{2d-2}} \\
\label{ix}
0&=&{(d-1)k \e^{2\sigma} \over r^2} 
+ {(d-1)\left(\sigma' - \rho' \right) \over r}
- {(d-1)(d-2) \over r^2} + \lambda^2\e^{2\sigma}\ .
\eea
If we introduce new variables $U$ and $V$ by
\be
\label{x}
U\equiv \e^{\rho + \sigma}\ ,\quad 
V\equiv r^{d-2}\e^{\rho - \sigma}\ , 
\ee
Eqs.(\ref{viii}), (\ref{ix}) and (\ref{vii}) are 
rewritten as follows
\bea
\label{xi}
0&=&(d-1)U'-{\alpha c^2 \over r V^2}U \\
\label{xii}
0&=&\left\{{(d-1)k \over r^2} + \lambda^2 \right\}U
- {(d-1) \over r^{d-1}}V' \\
\label{xiib}
\phi'&=&{c \over rV}
\eea
Deleting $U$ from (\ref{xi}) and (\ref{xii}), we obtain
\bea
\label{xiic}
0&=&V'' + \left[-{d-3 \over r} 
- {2\lambda^2 r \over (d-1)k + \lambda^2 r^2}\right]V' \nn
&&- {\alpha c^2V' \over (d-1)rV^2} \ .
\eea
When $c=0$ the solution is given by
\bea
\label{xviii}
U&=&1 \nn
V&=&V_0 \nn
&\equiv& {kr^{d-2} \over d-2} + {\lambda^2 \over d(d-1)}r^d 
- \mu\ .
\eea
Here $\mu$ corresponds to the mass of the black hole. $k=0$,
positive or negative corresponds to planar, spherical or
hyperbolic AdS BH, respectively. Using (\ref{xviii}), Eq.(\ref{xii}) 
and (\ref{xiic}) can be rewritten as follows:
\bea
\label{xviiib}
U&=&{V' \over V_0'} \nn
\label{xviiic}
0&=&\left({V' \over V_0'}\right)'-{\alpha c^2V' 
\over (d-1)rV_0'V^2}\ .
\eea

When $\mu=0$, the solution is isomorphic to AdS. If we choose 
$k<0$, the metric has the following form:
\be
\label{xix}
ds^2=-{(r^2 - r_0^2) \over l^2}dt^2 + {l^2 \over (r^2 - r_0^2) }dr^2 
+ r^2 \sum_{i,j=1}^{d-1}g_{ij}dx^i dx^j\ .
\ee
Here
\be
\label{xx}
l^2\equiv {d(d-1) \over \lambda^2}\ ,\quad 
r_0\equiv l \sqrt{-{k \over d-2}}\ .
\ee
The obtained AdS metric has a horizon at $r=r_0$. 
When $r\sim r_0$, the metric behaves as 
\be
\label{xxi}
ds^2 \sim -{2r_0 (r - r_0) \over l^2}dt^2 
+ {l^2 \over 2r_0 (r - r_0)}dr^2 + \cdots \ .
\ee
Then if we define a new coordinate $\rho$ by 
\be
\label{xxii}
\rho=l\sqrt{2(r-r_0) \over r_0}
\ee
the metric has the following form:
\be
\label{xxiii}
ds^2 \sim -{r_0 \over l^4}\rho^2 dt^2 + d\rho^2 + \cdots \ .
\ee
Therefore when we Wick-rotate $t$ by $t=i\tau$, $\tau$ has a 
period of ${2\pi l^2 \over r_0}$, whose inverse gives 
a temperature $T$: 
\be
\label{xxiv}
T={r_0 \over 2\pi l^2}={1 \over 2\pi l}\sqrt{-k \over d-2}\ .
\ee

We now consider the perturbation with respect to $c$. 
We will concentrate on the case of type IIB SG in $d=4$, 
by putting $\alpha={1 \over 2}$. Note that in this approximation 
the radius is away from horizon. Near-horizon regime will be 
discussed independently. 

For $\mu=0$ and $k<0$ case, the leading term for the dilaton $\phi$ 
is given by substituting $V_0$ in (\ref{xviii}) into (\ref{xiib})
\bea
\label{xxv}
\phi&=&\phi_0 +cl^2\left\{{1 \over 2r_0^4}\ln\left(1 
- {r_0^2 \over r^2}\right) + {1 \over 2r_0^2 r^2}\right\} \nn
&=& \phi_0 + c\left\{{1 \over 2l^6(2\pi T)^4}\ln\left(1 
- {l^4 (2\pi T)^2 \over r^2}\right) 
+ {1 \over 2l^2(2\pi T)^2 r^2}\right\}\ .
\eea
which gives the temperature dependent running dilaton. We should 
note that there is a singularity in the dilaton field at the horizon 
$r=r_0=2\pi l^2 T$. The fact that dilaton may become singular at 
IR has been mentioned already in two-boundaries AdS solution of 
IIB SG in ref.\cite{3}. It is also interesting that when $r$ is 
formally less than $r_0$ 
then dilaton (and also running coupling) 
becomes imaginary.

Since the string coupling is given by 
\be
\label{ci}
g=g_s\e^\phi\ \quad (g_s\ :\ \mbox{constant})\ ,
\ee
we find the behaviour when $r$ is large and $c$ is small as
\be
\label{cii}
g\sim g_s\left\{1 + cl^2\left(-{1 \over 2r^4 }
- {\left(2\pi l^2 T\right)^2 \over 3r^6} + {\cal O}
\left(r^{-8}\right) 
\right)
+ {\cal O}(c^2)\right\}\ .
\ee
Here $\phi_0$ has been absorbed into the redefinition of $g_s$. 
Since $r$ is the length scale corresponding to the radius of the 
boundary manifold, $r$ can be regarded as the energy scale of the 
field theory on the boundary \cite{10}. Therefore the 
beta-function is given by
\be
\label{ciii}
\beta(g)=r{dg \over dr}=-4\left(g-g_s\right)
+ {2^{5 \over 2} \over3}\left(2\pi T\right)^2l^3 g_s 
\left( {g_s - g \over c g_s} \right)^{3 \over 2}\ .
\ee
The first term is usual and universal \cite{4,7}. 
The second term defines the temperature dependence.
 
Let us comment on the case of high $T$. As we consider the 
behavior near the boundary, first we take $r$ to be large. After 
that we consider the case of high $T$. In this case $r\gg Tl^2$ 
and we can consider the large $T$ case in the expression 
(\ref{ciii}). The problem might happen when $r\sim Tl^2$. In this 
case, we need to solve Eq.(\ref{xxv}) with respect to $r$ as a 
function of $T$ and $\phi$ or coupling: $r=r(g,T)$. 
Then from (\ref{xxv}) and (\ref{ci}), we find the following 
expression of the beta-function:
\be
\label{gTii}
\beta(g) \sim \left.r{dg \over dr}\right|_{r=r(g,T)}
={g_s c l^2 \over r(g,T)^4 
\left(1 - {l^4 (2\pi T)^2 \over r(g,T)^2}\right) }\ .
\ee
In case $r$ is large, the above equation reproduces (\ref{ciii}).
We can also consider the case that the last term 
in (\ref{xxv}) is larger than the second term which contains 
$\ln (\cdots)$. In this case, the coupling is given by
\be
\label{gTib}
g\sim g_s\left( 1 + {c \over 2l^2 \left(2\pi T\right)^2 r^2}
\cdots \right)\ ,
\ee
which changes the leading behavior of the beta-function:
\be
\label{gTiib}
\beta(g)\sim - 2 \left(g-g_s\right) + \cdots\ .
\ee
This beta-function presumably defines strong coupling regime
of non-SUSY gauge theory at high temperature.
It is interesting to note that in perturbative gauge theory 
at non-zero 
temperature the running gauge coupling contains not only 
standard logarithms of $T$ but also terms linear on $T$  
(see ref.\cite{volodya} and references therein).
Of course, in our case we have not AF theory but the one 
with stable fixed point.

Now we consider the correction for $V$ and $U$ , writing them 
in the following form:
\be
\label{xxvi}
V=V_0+c^2 v\ ,\quad U=1+c^2 u\ .
\ee
 Substituting (\ref{xxvi}) and neglecting the higher orders in 
$c^2$, we obtain
\bea
\label{xxvii}
u&=&{v' \over V_0'} \nn
\label{xxviii}
0&=&\left({v' \over V_0'}\right)'-{1 \over 6rV_0^2}\ .
\eea
With $\mu=0$ and $k<0$
in the above equations one gets, 
\bea
\label{xxix}
u&=&{4 \over 3 k^4 l^4}\left\{-{1 \over 2s^2} - {2 \over s} 
\right. \nn
&& \left. -3\ln \left(1 - {1 \over s}\right) - {1 \over (s-1)}
+ c_1 \right\} \\
\label{xxx}
v&=&{2 \over 3k^2 l^2}\left\{ -{1 \over 2}\left(3s^2 - 3s +1 
\right)\ln \left(1 - {1 \over s}\right)  \right. \nn
&& \left. -{3s \over 2} + {3 \over 4} -{1 \over 4s} 
+{c_1 \over 2}\left(s^2 - s\right) + c_2\right\}
\eea
Here 
\be
\label{xxxi}
s=-{2 r^2 \over kl^2}
\ee
and $c_1$ and $c_2$ are constants of the integration, which should 
vanish if we require $u$, $v\rightarrow 0$ when 
$r\rightarrow \infty$.

 From (\ref{xxix}) and (\ref{xxx}), we find that $U$ and $V$ or 
$\e^{2\rho}$ and $\e^{2\sigma}$ have the singularity at the 
unperturbative horizon corresponding to $s=1$. Eq.(\ref{xxv}) tells 
also that the dilaton field is also singular there. In other words, the 
expansion with respect to $c^2$ breaks down when $s\sim 0$. Therefore 
the singularity in $U$, $V$ would not be real one.
 
In order to investigate the behavior in near-horizon regime 
we assume that the radius of the horizon is large 
and use ${1 \over r}$ expansion:
\be
\label{r1}
V={r^4 \over l^2} + {kr^2 \over 2} + {a \over r^4} + 
{\cal O}\left(r^{-6}\right)\ .
\ee
We put the constant term to be zero  assuming that the black hole 
mass vanishes. The absence of ${1 \over r^2}$ term can be found from 
(\ref{xviiib}). Eq.(\ref{xviiib}) also tells that 
\be
\label{r2}
a={c^2l^2 \over 48}
\ee
and $\e^\phi$, $V$ and $U$ have the following forms:
\bea
\label{r3}
\e^\phi&=&\e^{\phi_0}\left(1 - {cl^2 \over 4 r^4} 
+ {\cal O}\left(r^{-6}\right)\right) \nn
V&=&{r^4 \over l^2} + {kr^2 \over 2} + {c^2l^2 \over 48 r^4} + 
{\cal O}\left(r^{-6}\right) \nn
U&=&1-{c^2l^4 \over 192 r^8} + {\cal O}\left(r^{-10}\right)\ .
\eea
 From the equation $V=0$ we find the position of the horizon 
\be
\label{r4}
r=r_h\equiv l\sqrt{-{k \over 2}}\left(1 
- {c^2 \over 6 k^4 l^4}\right)\ ,
\ee
which gives the correction to the temperature:
\be
\label{r5}
T={1 \over 2\pi l}\sqrt{-{k \over 2}} 
- {c^2\left(-{k \over 2}\right)^{-{7 \over 2}} 
\over 192 l^5}\ .
\ee

Let us turn now to the analysis of
 the potential between quark and anti-quark\cite{5}.
We evaluate the following Nambu-Goto action
\be
\label{rg5}
S={1 \over 2\pi}\int d\tau d\sigma \sqrt{\det\left(g^s_{\mu\nu}
\partial_\alpha x^\mu \partial_\beta x^\nu\right)}\ .
\ee
with the ``string'' metric $g^s_{\mu\nu}$, which 
could be given by multiplying a dilaton function $\e^\phi$ to 
the metric tensor in (\ref{ii}). 
We consider the static configuration $x^0=\tau$, 
$x^1\equiv x=\sigma$, $x^2=x^3=\cdots=x^{d-1}=0$ and $r=r(x)$. 
 Choose the coordinates on the boundary manifold so that the 
line given by $x^0=$constant, 
$x^1\equiv x$ and $x^2=x^3=\cdots=x^{d-1}=0$ is geodesic and 
$g_{11}=1$ on the line. 
Substituting the configuration into (\ref{rg5}), we find
\be
\label{rg7}
S={{\cal T} \over 2\pi}\int dx \e^\phi(r)\sqrt{U(r)V(r)\left(
{U(r) \over V(r)}\left(\partial_x r\right)^2 + 1
\right)}\ .
\ee
Here ${\cal T}$ is the length of the region of the definition 
of $\tau$ and we choose $\phi_0=0$ for simplicity.
The orbit of $r$ can be obtained by minimizing the action $S$ 
or solving the Euler-Lagrange equation 
${\delta S \over \delta r}- \partial_x\left({\delta S 
\over \delta\left(\partial_x r\right)}\right)=0$. 
The Euler-Lagrange equation tells that 
\be
\label{rg8}
E_0=\e^\phi(r)\sqrt{U(r)V(r) \over 
{U(r) \over V(r)}\left(\partial_x r\right)^2 + 1 }
\ee
is a constant. If we assume $r$ has a finite minimum $r_{\rm min}$, 
where $\partial_x r|_{r=r_{\rm min}}=0$, $E_0$ is given by
\be
\label{rg9b}
E_0=\e^{\phi(r_{\rm min})}\sqrt{U(r_{\rm min})V(r_{\rm min})} \ .
\ee
Introducing a parameter $t$, we parametrize $r$ by
\be
\label{rg9}
r=r_{\rm min}\cosh t\ .
\ee
Then we find
\bea
\label{rg10}
{dx \over dt}&=&
{l \over r_{\rm min}\cosh^2t\left(\cosh^2t + 1
\right)^{1 \over 2}} \nn
&& \times \left\{1 + {kl^2 \over 4r_{\rm min}^2}
{\cosh^4 t - \cosh^2 t -1 \over \left(\cosh^2 t + 1 
\right)\cosh^2 t }
 + {\cal O}\left(r_{\rm min}^{-4}\right) \right\}\ .
\eea
Taking $t\rightarrow +\infty$, we find the distance $L$ between 
"quark" and "anti-quark" 
\bea
\label{rg11}
L&=& {lA \over r_{\rm min}} 
+ {kl^3 B \over 4r_{\rm min}^3} + {\cal O}\left(
r_{\rm min}^{-5}\right) \\
A&\equiv& \int_{-\infty}^\infty {dt \over \cosh^2t
\left(\cosh^2t + 1\right)^{1 \over 2}} 
=1.19814... \nn
B&\equiv& \int_{-\infty}^\infty dt {\cosh^4t - \cosh^2t -1 
 \over \cosh^4t
\left(\cosh^2t + 1\right)^{3 \over 2}} 
=-0.162061... \ .\nonumber
\eea
As one sees the next-to-leading correction to distance 
depends on the curvature of space-time\cite{NO} or temperature.

Eq.(\ref{rg11}) can be solved with respect to $r_{\rm min}$ 
and we find
\be
\label{rg12}
r_{\rm min}={lA \over L} + {klBL \over 4A^2} 
+ {\cal O}\left(L^3\right)\ .
\ee
Using (\ref{rg8}), (\ref{rg9}) and (\ref{rg11}), we find the 
following expression for the action $S$
\bea
\label{rg13}
S&=&{{\cal T} \over 2\pi}E(L) \\
E(L)&=&\int_{-\infty}^\infty dt {\cosh^2 t \over 
\left(\cosh^2 t + 1\right)^{1 \over 2}}\left\{1 + 
{kl^2 \over 4 r_{\rm min}^2} {1 \over \cosh^2 t 
\left(\cosh^2 t + 1\right)} 
+ {\cal O}\left(r_{\rm min}^{-4}\right)\right\}\ .
\nonumber 
\eea
Here $E(L)$  expresses the total energy of the 
``quark''-``anti-quark'' system.
The energy $E(L)$ in (\ref{rg13}), however, contains the divergence 
due to the self energies of the infinitely heavy ``quark'' 
and ``anti-quark''. 
The sum of their self energies can be estimated by considering the 
configuration $x^0=\tau$, $x^1=x^2=x^3=\cdots
=x^{d-1}=0$ and $r=r(\sigma)$ (note that $x_1$ vanishes here) 
and the minimum of $r$ is $r_D$, where branes would lie : 
$r_D\gg r_{\rm min}$. We devide the region for $r$ to two ones, 
$\infty>r>r_{\rm min}$ and $r_{\rm min}<r<r_D$. 
Using the parametrization of (\ref{rg9}) for the region 
$\infty>r>r_{\rm min}$, 
we find the following expression of the sum of self energies:
\bea
\label{rg14}
E_{\rm self}=2r_{\rm min}\int_0^\infty dt\, \sinh t
+ 2\left(r_{\rm min} - r_D \right) 
+ {\cal O}\left(r_{\rm min}^{-3}\right)\ .
\eea
Then the finite potential between ``quark'' and 
``anti-quark'' is given by
\bea
\label{rg15}
E_{q\bar q}(L)&\equiv&E(L) - E_{\rm self} \nn
&=&r_{\rm min}\left(C + {kl^2D \over 4r_{\rm min}^2}
+ {\cal O}\left(r_{\rm min}^{-4}\right)\right) \nn
&=&{lAC \over L} + {kl \over 4}\left({BC \over A^2} + {D \over A}
\right)L + {\cal O}\left(L^3\right) \\
&=&{lAC \over L} - {l^3 \left(2\pi T\right)^2 \over 2}
\left({BC \over A^2} + {D \over A}
\right)L + {\cal O}\left(L^3\right) \nn
C&=&2\int_0^\infty dt\,\left\{ 
{\cosh^2 t \over \left(\cosh^2 t + 1\right)^{1 \over 2}} 
-\sinh t\right\} -2 
=-1.19814... \nn
D&=& 2\int_0^\infty {dt 
\over \left(\cosh^2 t + 1\right)^{3 \over 2}}
=0.711959 \ .\nonumber
\eea
Here we neglected the $r_{\rm min}$ or $L$ independent term.
We should note that next-to-leading term is linear in $L$, which 
might be relevant to the confinement. For the confinement, it is 
necessary that the quark-antiquark potential behaves as 
\be
\label{cnfpt}
E_{q\bar q}\sim a L
\ee
with some positive constant $a$ for large $L$. For high temperature, 
it is usually expected that there occurs the phase transition to 
the deconfinement phase, where the potential behaves as Coulomb force, 
\be
\label{dcnfpt}
E_{q\bar q} \sim {a' \over L}\ .
\ee
Since ${BC \over A^2} + {D \over A}>0$ and 
$k<0$, the contribution from next-to-leading term in the potential 
is repulsive. The leading term expresses the repulsive but shows the 
Coulomb like behavior. The next-leading-term tells that the repulsive 
force is long-range than Coulomb force. 

The expression (\ref{rg15}) is correct even at high 
temperature if $L$ is small or $r_{\rm min}$ is large. If 
$r_{\rm min}$ is small and the orbit of string approaches to 
the horizon and/or enters inside 
the horizon, the expression would not be valid. Since the 
horizon is given by (\ref{xx}), the expression (\ref{rg15}) 
would be valid if
\be
\label{vali}
r_{\rm min}\gg r_0=l\sqrt{-{k \over 2}}
\ee
or  using (\ref{xxiv}) and (\ref{rg12}), 
\be
\label{valii}
L\ll A\sqrt{-{2 \over k}}=2\pi A l T\ .
\ee
The above condition (\ref{valii}) makes difficult to evaluate 
the potential quantitively by the analytic calculation when $L$ 
is large and numerical calculation would be necesssary. 
In order to investigate the qualitive behavior of the potential 
when $L$ is large, we consider the background where the dilaton 
is constant $\phi=\phi_0$, which would tell the effect of the 
horizon or finite temperature. As $c=0$ when the dilaton is 
constant, we can use the solution in (\ref{xviii}). Then by the 
calculation similar to (\ref{rg15}) but without assuming $L$ 
is small or $r_{\rm min}$ is large, we obtain the following 
expression of the quark-antiquark potential:
\bea
\label{PotlL}
E_{q\bar q}&=&r_{\rm min}
\int_{-\infty}^\infty dt \sinh t \left\{ \left( 1 - {1 \over 
\cosh^2 t}\cdot {1-{r_0^2 \over r_{\rm min}^2} \over 
\cosh^2 t - {r_0^2 \over r_{\rm min}^2}} \right)^{-{1 \over 2}}
-1 \right\} \nn
&&+ 2\left(r_D - r_{\rm min}\right)\ .
\eea
Constant $-1$ in $\{\ \}$ and the last term correspond to 
the subtraction of the 
self-energy. The integration in (\ref{PotlL}) converges and the 
integrand is monotonically decreasing function of 
${1 \over r_{\rm min}}$ if $r_{\rm min}$ is larger than the radius 
of the horizon $r_0$ : $r_{\rm min}>r_0$ and vanishes in the limit of 
$r_{\rm min}\rightarrow r_0$. Therefore if $r_{\rm min}$ 
decreases and approaches to $r_0$ when $L$ is large, which seems to 
be very natural, the potential $E_{q\bar q}$ approaches to a 
constant $E_{q\bar q}\rightarrow 2\left(r_D - r_0\right)$ 
and do not behaves as a linear function of $L$. This tells that the 
quark is not confined. This effect would corresponds to deconfining 
phase of QCD in the finite temperature.

We can also evaluate the potential between monopole and 
anti-monopole  using the Nambu-Goto action for $D$-string 
instead of (\ref{rg5}) (cf.ref.\cite{13}):
\be
\label{rg5m}
S={1 \over 2\pi}\int d\tau d\sigma \e^{-2\phi}
\sqrt{\det\left(g^s_{\mu\nu}
\partial_\alpha x^\mu \partial_\beta x^\nu\right)}\ .
\ee
For the static configuration $x^0=\tau$, $x^1\equiv x=\sigma$, 
$x^2=x^3=\cdots=x^{d-1}=0$ and $y=y(x)$,  we find, instead of 
(\ref{rg7})
\be
\label{rg7m}
S={{\cal T} \over 2\pi}\int dx \e^{-\phi(r)}\sqrt{U(r)V(r)\left(
{U(r) \over V(r)}\left(\partial_x r\right)^2 + 1
\right)}\ .
\ee
Since $\phi$ is proportional to $c$ and $V$ and $U$ contain $c$ in 
the form of its square $c^2$, the potential between monopole and 
anti-monopole is given by changing $c$ by $-c$ in the potential 
between quark and anti-quark. Since the expression (\ref{rg15}) 
does not contain $c$ in the given order, the potential 
$E_{m\bar m}(L)$ for monopole 
and anti-monopole is identical with that of quark and anti-quark in 
this order:
\be
\label{rg8m}
E_{m\bar m}(L)=E_{q\bar q}(L)\ .
\ee
Hence, we showed that non-constant dilaton deformation of IIB SG 
 vacuum changes the structure of potential and confinement 
is becoming non-realistic.

\section{Running coupling and quark-antiquark potential 
at finite temperature: non-zero mass BH case}

In this section we 
consider another interesting case that $k=0$ and 
$\mu\neq 0$, which corresponds to the throat limit of D3-brane 
\cite{GKT,GKP}\footnote{ In ref.\cite{GKT} $\alpha'$-corrections 
to leading term ($T^4$) of free energy for above AdS BH have 
been derived. The temperature was actually fixed. In the case 
under discussion we consider dilatonic deformation of such AdS BH 
using tree level bosonic sector of IIB SG. Thus, we define the 
corrections (next-to-the leading term on the temperature) to
solution (and free energy).}
and $V_0$ has the following form:
\be
\label{ki}
V_0={r^4 \over l^2} - \mu \ .
\ee
 $\e^{2\rho}$ and $\e^{2\sigma}$ have the following form:
\be
\label{kibb}
\e^{2\rho}=\e^{-2\sigma}={1 \over r^2}\left(
{r^4 \over l^2} - \mu\right)
\ee
Therefore when $c=0$, the horizon is given by \cite{GKT} 
\be
\label{kib}
r=\mu^{1 \over 4}l^{1 \over 2}
\ee
and the black hole temperature is 
\be
\label{kii}
T={\mu^{1 \over 4} \over \pi l^{3 \over 2}}\ .
\ee
In a way similar to $k<0$ and $\mu=0$ case, we obtain
\bea
\label{xxxii}
\phi&=&\phi_0 + {c \over 4\mu}\ln\left(1 - {1 \over q^2}\right) \nn
u&=&-{12 \over \mu^2}\left\{ {1 \over q^2 - 1} 
+ \ln\left(1 - {1 \over q^2}\right) + c_1'\right\} \nn
v&=&{1 \over 12\mu}\left\{ -q^2 \ln\left(1 
- {1 \over q^2}\right) - 1 + {c_1' q^2 \over 2} + c_2'\right\}\ .
\eea
Here
\be
\label{xxxiii}
q\equiv {r^2 \over l\sqrt\mu} \ 
\ee
and $c_1'$ and $c_2'$ are constants of the integration, which should 
vanish if we require $u$, $v\rightarrow 0$ when 
$r\rightarrow \infty$. The approximation when $r$ is far from 
horizon is again employed.

 Using (\ref{kii}) and (\ref{xxxii}), we find the behaviour 
of the string coupling (\ref{ci}) when $r$ is large and $c$ 
is small ($\phi_0$ is absorbed into the redefinition of $g_s$) :
\be
\label{civ}
g=g_s\left\{1 + {cl^2 \over 4}\left(-{1 \over r^4} 
- {\left(\pi T\right)^4 l^8 \over r^8} + {\cal O}
\left(r^{-12}\right)\right) + {\cal O}\left(c^2\right) 
\right\}\ .
\ee
The behavior of the second term is characteristic for $k=0$ case 
since the second term behaves as ${\cal O}\left(r^{-6}\right)$ 
for $k\neq 0$.\footnote{
The case that the boundary is the Einstein manifold with $k\neq 0$ 
has been discussed in \cite{NO}.} 
Eq.(\ref{civ}) gives the following beta-function 
\be
\label{cv}
\beta(g)=r{dg \over r}=-4 \left(g - g_s\right) 
+ {8 (\pi T )^4 l^6 \over g_s c} \left(g - g_s\right)^2
+ \cdots \ .
\ee
The first term is universal one \cite{4,7} but the behavior of 
the second temperature dependent term is characteristic for 
$k=0$. 

We now consider the high temperature and $r\sim Tl^2$ case. 
For this purpose, we write the coupling as follows:
\be
\label{hTi}
g=g_s\left(1 - {l^2\mu \over r^4}\right)^{c \over 4\mu}\ .
\ee
Here we used (\ref{xxxii}). Eq.(\ref{hTi}) can be solved with 
respect to ${1 \over r^4}$:
\be
\label{hTii}
{l^2\mu \over r^4}=1-\left( {g \over g_s} \right)^{4\mu \over c}\ .
\ee
On the other hand, Eq.(\ref{hTi}) gives 
\be
\label{hTiii}
r{dg \over dr}={g_s c l^2 \over r^4}\left(1 - {1 \over r^4}
\right)^{{c \over 4\mu} -1}\ .
\ee
Substituting (\ref{hTii}) into (\ref{hTiii}), we obtain the 
following expression:
\be
\label{hTiv}
\beta(g)=r{dg \over dr}
={gc \over \mu}\left( {g \over g_s} \right)^{-{4\mu \over c}}
\left( 1 - \left( {g \over g_s} \right)^{4\mu \over c}\right)\ .
\ee
Using (\ref{kii}), we find the following expression of the 
temperature dependent beta-function:
\be
\label{hTv}
\beta(g)={gc \over \pi^2 l^6T^4}
\left( {g \over g_s} \right)^{-{4\pi^2 l^6T^4 \over c}}
\left( 1 - \left( {g \over g_s} 
\right)^{4\pi^2 l^6T^4 \over c}\right)\ .
\ee
Since $T$ always appears in the combination of ${c \over T^4}$, 
the high temperature is consistent with the small $c$. As one can
see $T$-dependence is quite complicated. It is qualitatively different 
from the case of low temperature.

In order to investigate the corrections to the position of 
the horizon and the temperature, we use ${1 \over r}$ expansion 
as in the previous section assuming the
black hole is large. Then we find 
\bea
\label{rii1}
\e^\phi&=&\e^{\phi_0}\left(1 - {cl^2 \over 4 r^4} 
+ {\cal O}\left(r^{-8}\right)\right) \nn
U&=&1 - {c^2 l^4 \over 48 r^8} + {\cal O}\left(r^{-12}\right) \nn
V&=&{r^4 \over l^2} - \mu + {c^2 l^2 \over 48 r^4} 
+ {\cal O}\left(r^{-8}\right) \ .
\eea
Then corrections to the position of 
the horizon and the temperature are given as
\bea
\label{rii2}
r&=&r_h\equiv l^{1 \over 2}\mu^{1 \over 4}\left(1 
- {c^2 \over 192 \mu^2}\right) \nn
T&=&{\mu^{1 \over 4} \over \pi l^{3 \over 2}}\left(1 
- {5c^2 \over 192\mu^2}\right)\ .
\eea

The discussion of potential between quark and anti-quark 
for above case may be done similarly to the situation
when  $k<0$ and $\mu=0$. 
Instead of Eqs.(\ref{rg10}), (\ref{rg11}), (\ref{rg12}), 
we obtain
\bea
\label{Org10}
{dx \over dt}&=&
{l \over r_{\rm min}\cosh^2t\left(\cosh^2t + 1
\right)^{1 \over 2}}\nn
&& \times \left\{1 - {\mu l^2 \over 2r_{\rm min}^4}
{\cosh^4 t -1 \over \cosh^4 t}
- {c l^2 \over 4r_{\rm min}^4}
{\cosh^4 t + 1 \over \cosh^4 t}
 + {\cal O}\left(r_{\rm min}^{-8}\right) \right\}\\
\label{Org11}
L&=& {lA \over r_{\rm min}} 
- {\mu l^3 B_1 \over 2r_{\rm min}^5} 
- {c l^3 B_2 \over 4r_{\rm min}^5} 
+ {\cal O}\left(r_{\rm min}^{-9}\right) \\
B_1&\equiv& \int_{-\infty}^\infty dt 
{\sinh^2 t \left(\cosh^2t + 1\right)^{1 \over 2} \over \cosh^6t} 
= 0.479256... \nn
B_2&\equiv& \int_{-\infty}^\infty dt {\cosh^4t + 1 
 \over \cosh^6t\left(\cosh^2t + 1\right)^{1 \over 2}} 
= 1.91702... \\
\label{Org12}
r_{\rm min}&=&{lA \over L} - {\mu B_1 L^3 \over 2l A^4} 
- {c B_2 L^3 \over 4l A^4} + {\cal O}\left(L^7\right)\ .
\eea
Then we find the finite potential between ``quark'' and 
``anti-quark'' is given by
\bea
\label{Org15}
E_{q\bar q}(L)
&=&r_{\rm min}\left\{C 
+ {l^2 A \over r_{\rm min}^4}\left(
{\mu \over 2} - {5c \over 12}\right) 
+ {\cal O}\left(r_{\rm min}^8\right)\right\} \\
&=&{lAC \over L} + {L^3 \over lA^3}
\left\{\mu\left({A \over 2} - {C B_1 \over 2}
\right) + c \left(-{5A \over 12} - {C B_2 \over 8}
\right)\right\} + {\cal O}\left(L^7\right) \nn
&=&{lAC \over L} + {L^3 \over lA^3}
\left\{l^6\left(\pi T\right)^4
\left({A \over 2} - {C B_1 \over 2}
\right) + c \left(-{5A \over 12} - {C B_2 \over 8}
\right)\right\} \nn
&& + {\cal O}\left(L^7\right) \ .\nonumber
\eea
Here we choose $\phi_0=0$ and neglect $r_{\rm min}$ or $L$ 
independent terms, again.
The behavior of the potential is qualitatively identical with that 
in \cite{BISY} except $L^3$-term in potential (next-to-leading term) 
contains the contribution from dilaton. 
Since
\be
\label{TTi}
{A \over 2} - {CB_1 \over 2} = 0.886178... \ ,\quad 
-{A \over 4} - {CB_2 \over 2} = 0.649204... \ ,
\ee
the $L^3$ potential becomes attractive 
if $l^6\left(\pi T\right)^4 > \gamma c$ (high temperature 
or small dilaton) 
and repulsive 
if $l^6\left(\pi T\right)^4 < \gamma c$ (low temperature or 
large dilaton). Here
\be
\label{TTii}
\gamma \equiv {{5A \over 12} + {CB_2 \over 2} 
\over {A \over 2} - {CB_1 \over 2} }= -0.732589... \ .
\ee
Hence, we proved the possibility of confinement at finite 
temperature.
The potential (\ref{Org15}) is valid if $r_{\rm min}$ is 
much larger than the radius of the horizon:
\be
\label{TTiii}
r_{\rm min}\gg \mu^{1 \over 4}l^{1 \over 2}
\ee
or 
\be
\label{TTiv}
L\ll {Al^{1 \over 2} \over \mu^{1 \over 4}}=\pi l T\ .
\ee
The potential between monopole and anti-monopole is given by 
changing $c$ into $-c$ in (\ref{Org15}):
\bea
\label{Org15m}
E_{m\bar m}(L)
&=&{lAC \over L} + {L^3 \over lA^3}
\left\{l^6\left(\pi T\right)^4
\left({A \over 2} - {C B_1 \over 2}
\right) - c \left(-{5 \over 12} - {C B_2 \over 8}
\right)\right\} \nn
&& + {\cal O}\left(L^7\right) \ .\nonumber
\eea
Therefore 
the $L^3$ potential becomes attractive 
if $l^6\left(\pi T\right)^4 > -\gamma c$ 
and repulsive if $l^6\left(\pi T\right)^4 < -\gamma c$.
In other words, the behavior of monopole-antimonopole 
potential is reversed.

We now consider more general case where either  $k$ or $\mu$ 
do not vanish. If we define $r_\pm^2$ by
\be
\label{gi}
r_\pm^2\equiv {kl^2 \over 4}\left( -1 \pm 
\sqrt{1 + {16\mu \over k^2 l^2}}\right)\, 
\ee
$V_0$ has the following form:
\be
\label{gii}
V_0={1 \over l^2}\left( r^2 - r_+^2 \right)
\left( r^2 - r_-^2 \right)\ .
\ee
Since $\mu$ corresponds the black hole mass, $\mu$ should not be 
negative. 
If $\mu>0$, $r_+^2$ is positive and 
$r_-^2$ is negative when $k>0$ and $r_-^2$ is positive and 
$r_+^2$ is negative when $k<0$. 
Then $r=r_+$ corresponds to the horizon for $k>0$ 
in $c=0$ case and $r=r_-$ 
corresponds to the horizon for $k<0$. Therefore there is only 
one horizon when $\mu>0$. 
On the other hand, when $\mu<0$ although it might look unphysical, 
there are two horizons corresponding to $r=r_\pm$ when $k<0$.
 
When $c=0$, the temperature corresponding to the horizon at 
$r=r_\pm$ is given by 
\be
\label{giib}
T=\pm {r_+^2 - r_-^2 \over 2\pi r_\pm l^2}\ .
\ee

When $c$ is small but does not vanish, the leading behavior of 
$\phi$ is given by, 
\bea
\label{giii}
\phi&=& \phi_0 + {cl^2 \over 2}\left\{ 
- {1 \over r_+^2 r_-^2}\ln r^2 \right. \nn
&& \left. + {1 \over r_+^2 \left(r_+^2 - r_-^2 \right)}
\ln \left(r^2 - r_+^2 \right)
- {1 \over r_-^2 \left(r_+^2 - r_-^2 \right)}
\ln \left(r^2 - r_-^2 \right)\right\}\ .
\eea
Then the behavior 
of the string coupling (\ref{ci}) when $r$ is large and $c$ 
is small ($\phi_0$ is absorbed into the redefinition of $g_s$) :
\be
\label{cvi}
g=g_s\left\{1 + {cl^2 \over 2}\left(-{1 \over 2r^4} 
- {r_+^2 + r_-^2 \over 3r^6} + {\cal O}\left(r^{-8}\right)\right)
+ {\cal O}\left(c^2\right)\right\}\ ,
\ee
and the beta-function is given by 
\be
\label{cvii}
\beta(g)=r{dg \over dr}=-4\left(g-g_s\right)
+ {8 \over 3}\left(r_+^2 + r_-^2\right){cg_s \over l}
\left({g-g_s \over cg_s}\right)^{3 \over 2}\ .
\ee
Note that the second term vanishes when $k=0$, which is the reason 
why the behavior of the next-to-leading term in $k=0$ is different 
from that in $k\neq 0$. Eq.(\ref{giib}) gives the temperature 
dependence in the coupling (\ref{cvi}) and the beta-function 
(\ref{cvii}). The next-to-leading term shows again power-like 
behavior on  $g$
as it happened already in IIB SG solutions of refs.\cite{3,4,6,7,8,
9,NO}
(no temperature) and in GUTs with large internal dimensions
\cite{12}.
Note that two of the parameters $k$, $\mu$ and $T$ are 
independent. If we consider the high temperature regime  
by fixing $k$, $\mu$ becomes large and the behavior approaches 
to $k=0$ case in (\ref{hTiv}). On the other hand, if we consider 
the high temperature regime by fixing $\mu$, $k$ is positive and 
becomes large and the behavior approaches 
to $k<0$ and $\mu=0$ case in (\ref{gTii}). 
One can also find quark-antiquark potential which looks very complicated 
so we do not write it explicitly.

The corrections to $U$ and $V$ coming from the non-trivial 
dilaton are given by
\bea
\label{giv}
u&=&c_1'' + {l^4 \over \left(r_+^2 - r_-^2 \right)}
\left\{ \left({1 \over r_+^2} - {1 \over r_-^2} \right)^2
\ln r^2 \right. \nn
&& - {1 \over r_+^2}{1 \over r^2 - r_+^2} 
- {3r_+^2 - r_-^2 \over r_+^4 \left(r_+^2 - r_-^2 \right)}
\ln \left( r^2 - r_+^2 \right) \nn
&& \left. - {1 \over r_-^2}{1 \over r^2 - r_-^2} 
+ {3r_-^2 - r_+^2 \over r_-^4 \left(r_+^2 - r_-^2 \right)}
\ln \left( r^2 - r_-^2 \right) \right\} \\
\label{gv}
v&=&{1 \over l^2}\left[ c_2'' 
+ c_1'' \left\{r^4 - \left(r_+^2 + r_-^2 \right)r^2\right\} 
+ {l^4 \over \left(r_+^2 - r_-^2 \right)^2 }\left\{
- \left({1 \over r_+^2} + {1 \over r_-^2} \right) r^2 
\right.\right. \nn 
&& + \left( -{3r_+^2 - r_-^2 \over 
r_+^4\left(r_+^2 - r_-^2 \right)}\left(r^2 - r_+^2 \right)^2 
\right. \nn
&& \left. - {3r_+^2 - r_-^2 \over r_+^4}\left(r^2 - r_+^2 \right)
- {r_+^2 - r_-^2 \over r_+^2}\right)
\ln \left(1 - {r_+^2 \over r^2} \right) \nn
&& + \left( {3r_-^2 - r_+^2 \over 
r_-^4\left(r_+^2 - r_-^2 \right)}\left(r^2 - r_-^2 \right)^2
\right. \nn
&& \left.\left.\left. - {3r_-^2 - r_+^2 \over r_-^4}\left(r^2 - r_-^2 \right)
+ {r_+^2 - r_-^2 \over r_-^2}\right)
\ln \left(1 - {r_-^2 \over r^2}\right) 
\right\} \right] \ .
\eea
Here $c_1''$ and $c_2''$ are constants of the integration, 
which should vanish if we require $u$, $v\rightarrow 0$ when 
$r\rightarrow \infty$.

\section{Thermodynamics of approximate AdS backgrounds of IIB 
supergravity}

In the present section we will be interesting in 
the thermodynamical quantities like free energy.
After Wick-rotating the time variables by $t\rightarrow i\tau$, 
the free energy $F$ can be obtained from the action $S$ in (\ref{i}) 
where the classical solution is substituted:
\be
\label{F1}
F={1 \over T}S\ .
\ee
Multiplying $G^{\mu\nu}$ with the equation of motion in 
(\ref{iit}), we find
\be
\label{F2}
R - {1 \over 2} G^{\mu\nu}\partial_\mu \phi \partial_\nu \phi 
={5 \over 3}\Lambda\ .
\ee
Here we only consider the case of $d=4$ and $\alpha={1 \over 2}$. 
Substituting (\ref{F2}) into (\ref{i}), we find after the 
Wick-rotation
\be
\label{F3}
S={1 \over 2\pi G l^2}\int d^5 x \sqrt{G}\ .
\ee
Here we used (\ref{xx}). From the expressions of the metric 
$G_{\mu\nu}$ in (\ref{ii}) and (\ref{x}), Eq.(\ref{F3}) is rewritten 
as follows:
\be
\label{F4}
S={1 \over 2\pi G l^2}{V_3 \over T}\int_{r_h}^\infty 
dr r^3 U\ .
\ee
Here $V_3$ is the volume of 3d Einstein manifold and 
$r_h$ is the radius of the horizon and we assume 
$\tau$ has a period of ${1 \over T}$. 
Since $U$ has a singularity at $r=r_h$ in the expansion 
with respect to $c$, we use ${1 \over r}$ expansion. 
Furthermore the expression of $S$ contains the divergence 
coming from large $r$. In order to subtract the divergence, we 
regularize $S$ in (\ref{F4}) by cutting off the integral at 
a large radius $r_{\rm max}$. After that we subtract the 
divergent part.
In case of $k<0$ and $\mu=0$, we subtract it by using 
the extremal solution with $c=0$ ($U=1$): 
\be
\label{F4b}
S_{\rm reg}={1 \over 2\pi G l^2}{V_3 \over T}\left(
\int_{r_h}^{r_{\rm max}} 
dr r^3 U - \sqrt{V(r=r_{\rm max}) \over V^{\rm ex}(r=r_{\rm max}) }
\int_{r^{\rm ex}_h}^{r_{\rm max}} dr r^3 \right)  \ .
\ee
Here 
\be
\label{F4c}
V^{\rm ex}\equiv {1 \over l^2}\left(r^2 - {r^{\rm ex}_h}^2
\right)^2\ ,\quad r^{\rm ex}_h={l\sqrt{-k} \over 2}\ ,
\ee
which corresponds to the extremal solution (the solution has 
negative mass parameter $\mu=-{k^2 l^2 \over 4}$). 
The factor $\sqrt{V(r=r_{\rm max}) 
\over V^{\rm ex}(r=r_{\rm max}) }$ is chosen so that the proper 
length of the circle which corresponds to the period ${1 \over T}$ 
in the Euclidean time at $r=r_{max}$ coincides 
with each other in the two solutions with $\mu=0$ and $k<0$ one 
and extremal one in (\ref{F4c}). 
Then we obtain 
\be
\label{F5}
F={V_3 \over 2\pi G l^2}\left(-{5k^2 l^4 \over 128} 
+ {c^2 \over 48 k^2}\right)\ .
\ee
With the help of (\ref{r5}), we find the following expression
\be
\label{F6}
F=-{V_3 \over 2\pi G l^2}\left(
{ 5l^8\left(2\pi T\right)^4 \over 32} 
+ {c^2 \over 768 l^4 \left(2\pi T\right)^4}\right)\ .
\ee
In order to get the entropy ${\cal S}$, 
we need to know $T$ dependence of $V_3$ although $V_3$ is infinite. 
Since $k$ is proportional 
the curvature, $V_3$ would be proportional to $k^{-{3 \over 2}}$. 
Then we find
\bea
\label{F7}
{dV_3 \over dT}&=&{1 \over k}{dk \over dT}k{dV_3 \over dk} \nn
&=&-{3V_3 \over 2T}\left(1 - {c^2 \over 6l^{12} 
\left(2\pi T\right)^8} + \cdots \right)\ .
\eea
Therefore the entropy ${\cal S}$ and mass (energy) $E$ are given by
\bea
\label{F8}
{\cal S}&=&-{dF \over dT}={V_3 \over 2\pi G l^2 T }\left(
{ 25l^8\left(2\pi T\right)^4 \over 64} 
+ {49c^2 \over 1536 l^4 \left(2\pi T\right)^4}\right) \nn
E&=&F+T{\cal S}={V_3 \over 2\pi G l^2}\left(
{ 15 l^8\left(2\pi T\right)^4 \over 64} 
+ {47 c^2 \over 1536 l^4 \left(2\pi T\right)^4}\right)\ .
\eea
In terms of string theory correspondence\cite{GKT}, 
the parameters $G$ and $l$ 
are given by
\bea
\label{spara}
l^4&=&2g_{YM}^2 N{\alpha'}^2 \nn
Gl&=&{\pi g_{YM}^2{\alpha'}^2 \over N}\ .
\eea
Here the Yang-Mills coupling $g_{YM}$ is given by the string 
coupling $g_s$: $g_{YM}^2=2\pi g_s$ and $N$ is the number of the 
coincident D3-branes.
As $V_3$ is now dimensionless, we multiply $l^3$ with $V_3$:
\be
\label{F10}
\tilde V_3\equiv l^3 V_3\ .
\ee
Then Eqs.(\ref{F6}) and (\ref{F8}) can be rewritten as follows:
\bea
\label{F10b}
F&=&-{\tilde V_3 \over 2\pi^2 }\left(
{ 5N^2 \left(2\pi T\right)^4 \over 16} 
+ {c^2 \over 3072 l^4 g_{YM}^6 N{\alpha'}^6
\left(2\pi T\right)^4}\right)\nn
{\cal S}&=&{V_3 \over 2\pi^2 T }\left(
{ 25N^2\left(2\pi T\right)^4 \over 32} 
+ {49 c^2 \over 6144 g_{YM}^6 N{\alpha'}^6
\left(2\pi T\right)^4}\right) \nn
E&=&{V_3 \over 2\pi G l^2}\left(
{ 15 N^2\left(2\pi T\right)^4 \over 32} 
+ {47 c^2 \over 6144 g_{YM}^6 N{\alpha'}^6
\left(2\pi T\right)^4}\right)\ .
\eea

For $k=0$ and $\mu>0$ case, we can obtain the thermodynamical 
quantities in a similar way  using Eqs. (\ref{rii1}) and 
(\ref{rii2}) in ${1 \over r}$ expansion. 
We regularize $S$ in (\ref{F4}) by subtracting the solution 
with $\mu=0$ and $c=0$ ($U=1$):
\be
\label{F4bb}
S_{\rm reg}={1 \over 2\pi G l^2}{V_3 \over T}\left(
\int_{r_h}^{r_{\rm max}} 
dr r^3 U - 
\sqrt{V(r=r_{\rm max}) \over V(r=r_{\rm max}, \mu=0) }
\int_0^{r_{\rm max}} dr r^3 \right) \ .
\ee
We can assume here that $V_3$ 
does not depend on $T$ since $k$ is fixed to vanish. 
Then we obtain for the case
\bea
\label{F9}
F&=&-{V_3 \over 4\pi G l^2}\left(
{ l^8\left(\pi T\right)^4 \over 4} 
+ {5c^2 \over 192 l^4 \left(\pi T\right)^4}\right) \nn 
{\cal S}&=&{V_3 \over 4\pi G l^2 T}\left(
l^8\left(\pi T\right)^4 
- {5c^2 \over 48 l^4 \left(\pi T\right)^4}\right)\nn 
E&=&{V_3 \over 4\pi G l^2}\left(
{ 3l^8\left(\pi T\right)^4 \over 4} 
- {25c^2 \over 192 l^4 \left(\pi T\right)^4}\right)\ .
\eea
Note that the leading term in ${\cal S}$ is the volume of 
3d manifold at horizon (${V_3 r_h^3 \over l^3}$) divided by 
$4G$. 
Then by using (\ref{spara}) and (\ref{F10}), we find 
\bea
\label{F11}
F&=&-{\tilde V_3 \over 4\pi^2}\left(
{ N^2 \left(\pi T\right)^4 \over 2} 
+ {5c^2 \over 768 g_{YM}^6 N{\alpha'}^6 \left(\pi T\right)^4}\right) \nn 
{\cal S}&=&{\tilde V_3 \over 4\pi^2 T}\left(
2 N^2 \left(\pi T\right)^4 
- {5c^2 \over 192 g_{YM}^6 N{\alpha'}^6 \left(\pi T\right)^4}\right)\nn 
E&=&{\tilde V_3 \over 4\pi^2 }\left(
{ 3 N^2 \left(\pi T\right)^4 \over 2} 
- {25c^2 \over 768 g_{YM}^6 N{\alpha'}^6
 \left(\pi T\right)^4}\right)\ .
\eea
The leading behaviours of $F$ and $S$ are  consistent with \cite{GKT}.
As we used ${1 \over r}$ expansion, the second terms  
in (\ref{F11}) become 
dominant when the radius of horizon $r_h$ is large 
and the parameter $c$ specifying non-trivial dilaton is not very small. 
Notice that in other temperature regimes (or using another 
schemes for approximated solutions of gravitational equations) 
one can get also qualitatively different thermodynamical 
next-to-leading terms (on temperature).
One has to remark that leading term in above free energy describes the 
strong coupling regime free energy for ${\cal N}=4$ 
super YM theory with 
the usual mismatch factor 3/4 if compare with perturbative free energy
(for a detailed discussion of this case, see\cite{GKT}).

\section{Discussion}

We studied the approximate (dilaton perturbed) solutions 
of IIB SG near BH-like ${\rm AdS}_5\times{\rm S}_5$ background. 
Thanks to presence of dilaton, the running gauge coupling 
of non-SUSY gauge theory at finite temperature may be 
extracted from these solutions. It is interesting that 
corresponding strong coupling regime beta-function 
may depend on the temperature in the complicated way (mainly, 
power-like behavior). We also estimated quark-antiquark 
(and monopole-antimonopole) potential at finite temperature 
from SG side. Its comparison with the potential of 
${\cal N}=4$ super YM theory at finite temperature is also 
done. It is remarkable that confinement depending on features 
of geometry and dilaton may occur.

As our IIB SG solutions are approximate (actually large 
radius expansion)  it is clear that one is able to develop 
other schemes to search for similar solutions. Unfortunately, 
it is not yet clear how to identify explicitly boundary non-SUSY 
thermal gauge theory corresponding to these solutions. One 
possibility is to calculate free energy from SG side 
(with non-trivial dilaton) and compare it with free energy of 
perturbative thermal gauge theories with running gauge coupling.
The last quantity is available in QFT. 

Note also that one can generalize our solutions via adding
RR-scalar(axion) to 
bosonic sector of IIB SG as it was discussed in refs.\cite{14}.
As it follows from results of refs.\cite{7,NO} the structure 
of running gauge coupling changes drastically in this case. 
In particular, part of supersymmetries may be unbroken \cite{14}
 but asymptotic freedom may be realized in strong 
coupling phase \cite{7,NO}. We expect that at finite temperature 
this property may survive.

\ 

\noindent
{\bf Acknoweledgements} We thank J. Ambj\o rn for the interest to this work.

\end{document}